\documentclass[preprint,aps,prc,tightenlines,nofootinbib,showpacs]{revtex4}
\begin{document}
\input{psfig}
\input{epsf}
\def\Im{\mbox{\sl Im\ }}
\def\pd{\partial}
\def\oln{\overline}
\def\olft{\overleftarrow}
\def\ds{\displaystyle}
\def\bgreek#1{\mbox{\boldmath $#1$ \unboldmath}}
\def\sla#1{\slash \hspace{-2.5mm} #1}
\newcommand{\bra}{\langle}
\newcommand{\ket}{\rangle}
\newcommand{\vep}{\varepsilon}
\newcommand{\met}{{\mbox{\scriptsize met}}}
\newcommand{\lab}{{\mbox{\scriptsize lab}}}
\newcommand{\cm}{{\mbox{\scriptsize cm}}}
\newcommand{\mcal}{\mathcal}
\newcommand{\Del}{$\Delta$}
\newcommand{\g}{{\rm g}}
\long\def\Omit#1{}
\long\def\omit#1{\small #1}
\def\beq{\begin{equation}}
\def\eeq{\end{equation} }
\def\bea{\begin{eqnarray}}
\def\eea{\end{eqnarray}}
\def\eqref#1{Eq.~(\ref{eq:#1})}
\def\eqlab#1{\label{eq:#1}}
\def\figref#1{Fig.~\ref{fig:#1}}
\def\figlab#1{\label{fig:#1}}
\def\tabref#1{Table \ref{tab:#1}}
\def\tablab#1{\label{tab:#1}}
\def\secref#1{Section~\ref{sec:#1}}
\def\seclab#1{\label{sec:#1}}
\def\VYP#1#2#3{{\bf #1}, #3 (#2)}  
\def\NP#1#2#3{Nucl.~Phys.~\VYP{#1}{#2}{#3}}
\def\NPA#1#2#3{Nucl.~Phys.~A~\VYP{#1}{#2}{#3}}
\def\NPB#1#2#3{Nucl.~Phys.~B~\VYP{#1}{#2}{#3}}
\def\PL#1#2#3{Phys.~Lett.~\VYP{#1}{#2}{#3}}
\def\PLB#1#2#3{Phys.~Lett.~B~\VYP{#1}{#2}{#3}}
\def\PR#1#2#3{Phys.~Rev.~\VYP{#1}{#2}{#3}}
\def\PRC#1#2#3{Phys.~Rev.~C~\VYP{#1}{#2}{#3}}
\def\PRD#1#2#3{Phys.~Rev.~D~\VYP{#1}{#2}{#3}}
\def\PRL#1#2#3{Phys.~Rev.~Lett.~\VYP{#1}{#2}{#3}}
\def\FBS#1#2#3{Few-Body~Sys.~\VYP{#1}{#2}{#3}}
\def\AP#1#2#3{Ann.~of Phys.~\VYP{#1}{#2}{#3}}
\def\ZP#1#2#3{Z.\ Phys.\  \VYP{#1}{#2}{#3}}
\def\ZPA#1#2#3{Z.\ Phys.\ A\VYP{#1}{#2}{#3}}
\def\half{\mbox{\small{$\frac{1}{2}$}}}
\def\quarter{\mbox{\small{$\frac{1}{4}$}}}
\def\nn{\nonumber}
\newlength{\PicSize}
\newlength{\FormulaWidth}
\newlength{\DiagramWidth}
\newcommand{\vslash}[1]{#1 \hspace{-0.42 em} /}
\newcommand{\qslash}[1]{#1 \hspace{-0.46 em} /}
\def\her{\marginpar{$\Longleftarrow$}}
\def\bel{\marginpar{$\Downarrow$}}
\def\abo{\marginpar{$\Uparrow$}}



\title{$\Delta$ resonance contribution to two-photon exchange in
electron-proton scattering}

\author{S.~Kondratyuk}
\affiliation{Department of Physics and Astronomy, University of Manitoba,
Winnipeg, MB, Canada R3T 2N2}
\author{P.~G.~Blunden}
\affiliation{Department of Physics and Astronomy, University of Manitoba,
Winnipeg, MB, Canada R3T 2N2}
\author{W.~Melnitchouk}
\affiliation{Jefferson Lab, 12000 Jefferson Avenue, Newport News, VA 23606, USA}
\author{J.~A.~Tjon}
\affiliation{Jefferson Lab, 12000 Jefferson Avenue, Newport News, VA 23606, USA}
\affiliation{Department of Physics, University of Maryland, College Park, MD 20742-4111, USA}

\date{\today}

\begin{abstract}
We calculate the effects on the elastic electron-proton scattering cross section of the
two-photon exchange contribution with an intermediate $\Delta$ resonance.
The $\Delta$ two-photon exchange contribution is found to be smaller in magnitude
than the previously evaluated nucleon contribution, with an opposite sign at
backward scattering angles. The sum of the nucleon and $\Delta$ two-photon
exchange corrections has an angular dependence compatible with both the polarisation transfer and
the Rosenbluth methods of measuring the nucleon electromagnetic form factors.
\end{abstract}

\maketitle



The electromagnetic form factors reflect the essentially non-local nature of the nucleon in its interactions with photons.
As the basic observables parametrising nucleon compositeness, 
the form factors have long been studied both experimentally and theoretically. 
This interest has been renewed recently due to the increased precision of 
electron-proton scattering experiments and the availability of two alternative methods of extracting the form factors from the data:
the Rosenbluth method -- also known as the longitudinal-transverse (LT) separation 
technique~\cite{Wal94, Qat05} -- and the polarisation-transfer (PT) technique~\cite{Jon00}. 
If one uses the traditional one-photon exchange calculation to extract the form factors, the two methods
lead to apparently incompatible results: while the PT method yields a ratio
of the electric to magnetic form factors which falls off linearly with the square of the momentum transfer $Q^2$, 
the LT separation experiments give an approximately constant ratio~\cite{Jon00,Bra02,Arr03}. 
Finding an explanation of this discrepancy is important for the use of electron-proton scattering as a precise and reliable tool in hadronic physics.
  
Several theoretical studies~\cite{Blu03, Gui03} have suggested that
the problem could be at least partially resolved by including higher-order two-photon exchange corrections in the
analysis of electron-proton scattering data, in addition to the lowest order one-photon exchange (Born) approximation. 
The recent explicit calculation~\cite{Blu03} has shown that with the 
two-photon exchange taken into account in the analysis of electron-proton scattering, 
the ratio of the form factors extracted from the LT separation measurements 
becomes more compatible with the ratio from the PT experiments. However, the two-photon exchange diagrams calculated in Ref.~\cite{Blu03} contained only 
nucleons in the intermediate state; the contribution of other
hadrons has not been included until now. 
In view of the prominent role of the $\Delta$ resonance (unlike other excited states) in many
hadronic reactions, it is essential to evaluate its contribution to the two-photon exchange in
electron-proton scattering. Without an explicit calculation the results with only the nucleon intermediate state can only be viewed as suggestive in resolving the
discrepancy. Some aspects of the $\Delta$ contribution 
were addressed before~\cite{Dre57, Cam69}, using various approximate approaches.
These earlier studies demonstrated the importance of treating the $\Delta$ on a par with the nucleon in considering higher-order corrections to electron-proton scattering.

This letter presents a quantum field theoretical calculation of the two-photon exchange ``box" and ``crossed-box" diagrams with a $\Delta$ resonance in the intermediate state. 
We will show that the $\Delta$ two-photon
exchange correction is somewhat smaller in magnitude than that of the nucleon. 
At backward scattering angles the $\Delta$ and nucleon contributions tend to
partially cancel each other, their sum nevertheless yielding a predominantly negative 
two-photon exchange correction.
We will show that the modified cross section 
has an angular dependence consistent with both the LT separation and PT measurements of the form factors.


We consider scattering of electrons (mass $m_e \approx 0.511 \times 10^{-3}$ GeV) off protons 
(mass $M_N \approx 0.938$ GeV) with the
four-momenta assigned as $ e (p_1) + p (p_2) \rightarrow e (p_3) + p (p_4) $.
The differential cross section for this process is written in the form
$ d \sigma = d \sigma_B (1+\delta_N + \delta_\Delta) $
where $d \sigma_B$ is the lowest-order Born contribution (i.~e.~the cross section obtained from the
one-photon exchange tree diagram) and $\delta_N$ ($\delta_\Delta$) is the 
higher-order correction obtained from two-photon exchange diagrams 
containing nucleons ($\Delta$'s) in the intermediate state.
(Other higher-order effects which should be included in the
formula for $d \sigma$ -- such as the vacuum polarisation and the electron-photon vertex corrections -- have been extensively studied in the past and are known~\cite{Max00} to be irrelevant to the differences between the PT and LT analyses; we therefore focus here on 
the two-photon exchange effects only.)
It is convenient to divide $d \sigma$ by the 
well-known factor describing the scattering from a structureless ``proton" 
(see, e.~g.,~\cite{Bjo64}) and thus use the reduced cross section 
\beq
d \sigma_R = \left[\, G_M^2(Q^2) + {\epsilon \over \tau} G_E^2(Q^2)\, \right] (1+\delta_N +\delta_\Delta)\,.
\eqlab{xsec_red}
\eeq
Here the Born contribution is written in terms of the electric and magnetic form 
factors of the proton,
$G_E(Q^2)$ and $G_M(Q^2)$, which are functions of the momentum transfer squared
$Q^2 \equiv -q^2 \equiv 4 \tau M_N^2  = -(p_1-p_3)^2 $. The kinematic variable $\epsilon$ is related
to the scattering angle $ \theta $ through $ \epsilon= [ 1+2( 1+\tau ) \tan^2(\theta/2) ]^{-1} $,
which is equal to the photon polarisation in the Born approximation.

We denote the Born scattering amplitude as 
${\mathcal{M}}_B$ and the two-photon exchange amplitudes with the
nucleon and $\Delta$ intermediate states as
${\mathcal{M}}_N^{\gamma \gamma}$ and
${\mathcal{M}}_\Delta^{\gamma \gamma}$, respectively. From the equation
$d \sigma = d \sigma_B (1+\delta_N + \delta_\Delta) =
\left| {\mathcal{M}}_B + {\mathcal{M}}_N^{\gamma \gamma} + 
{\mathcal{M}}_\Delta^{\gamma \gamma} \right|^2$, 
where $d \sigma_B = \left| {\mathcal{M}}_B \right|^2$, we derive
to first order in the electromagnetic coupling $e^2/(4\pi) \approx 1/137$: 
\beq
\delta_{N, \Delta} = 
2 \frac{\mbox{Re} \left({\mathcal{M}}_B^\dagger \, {\mathcal{M}}_{N, \Delta}^{\gamma
\gamma} \right)}{ \left| {\mathcal{M}}_B \right|^2 }\,.
\eqlab{del_del}
\eeq
The nucleon part $\delta_N$ of the two-photon exchange was analysed in Ref.~\cite{Blu03}.
Below we will evaluate the $\Delta$ two-photon exchange contribution $\delta_\Delta$. The
scattering amplitude $ {\mathcal{M}}_{\Delta}^{\gamma \gamma} $ is given by the sum of the box
and crossed-box loop diagrams depicted in~\figref{boxdel}.
\begin{figure}[!htb]
\centerline{{\epsfxsize 7.7cm \epsffile[15 330 570 470]{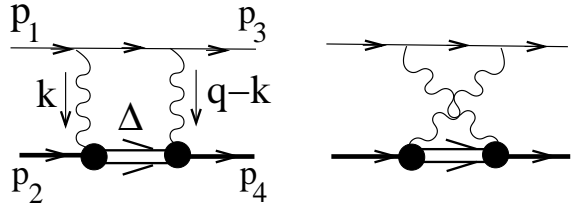}}}
\caption[f1]{Two-photon exchange box and crossed-box graphs for electron-proton scattering
with a $\Delta$ intermediate state, calculated in the present letter.
\figlab{boxdel}}
\end{figure}


We use the $\gamma N \Delta$ vertex of the following form~\cite{Kon01}:
\bea
&\Gamma_{\gamma \Delta \rightarrow N}^{\nu \alpha}(p,q) 
 \equiv   i V^{\nu \alpha}_{\Delta in}(p,q) = i {\ds \frac{e F_\Delta(q^2)}{2 M_{\Delta}^2}} \bigg\{
g_1 \left[\, g^{\nu \alpha} \vslash{p} \qslash{q} - p^\nu \gamma^\alpha \qslash{q} 
- \gamma^\nu \gamma^\alpha p \cdot q + \gamma^\nu \vslash{p} q^\alpha\, \right]  & \nn \\
& + g_2 \left[\, p^\nu q^\alpha - g^{\nu \alpha} p \cdot q\, \right] 
 + (g_3/M_{\Delta}) \left[\,q^2 (p^\nu \gamma^\alpha - g^{\nu \alpha} \vslash{p}) +
q^\nu (q^\alpha \vslash{p} - \gamma^\alpha p \cdot q )\, \right] \bigg\} \gamma_5\, T_3\,, &
\eqlab{vert}
\eea
where $M_{\Delta} \approx 1.232$ GeV is the $\Delta$ mass, $p_{\alpha}$ and $q_{\nu}$ are the 
four-momenta of the incoming $\Delta$ and photon, respectively, and
$g_1$, $g_2$ and $g_3$ are the coupling constants.\footnote{We use
the notation and conventions of Ref.~\cite{Bjo64} throughout.}
An analysis of~\eqref{vert} in the $\Delta$ rest frame
suggests that $g_1$, $g_2-g_1$ and $g_3$ may be interpreted as magnetic, electric 
and Coulomb components, respectively, of the $\gamma N \Delta$ vertex. 
The form factor in~\eqref{vert} is necessary for ultraviolet regularisation of the
loop integrals evaluated below; we use the simple dipole form
\beq
F_\Delta(q^2) = \frac{\Lambda_\Delta^{4}}{\left(\Lambda_\Delta^2 - q^2\right)^2}\,,
\eqlab{ff}
\eeq
where $\Lambda_\Delta$ is the cutoff.
The form factor entails some model-dependence of our results, which is unavoidable in any dynamical hadronic calculation.
The isospin transition operator $T_3$ is defined by the relations
$\sum_{\alpha=1}^3 T_\alpha^\dagger T_\alpha = 1 $ and
$T_\alpha T_\beta^\dagger = \delta_{\alpha \beta} - {\tau_\alpha \tau_\beta}/{3}$,
where $\tau_{1,2,3}$ are the usual Pauli matrices.
The vertex with an outgoing $\Delta$ is given by the Dirac conjugate of \eqref{vert},
$\Gamma_{\gamma N \rightarrow \Delta}^{\alpha \nu}(p,q) 
\equiv i V^{\alpha \nu}_{\Delta out}(p,q) =
\gamma_0 \left[ \Gamma_{\gamma \Delta \rightarrow N}^{\nu \alpha}(p,q) \right]^\dagger 
\!\gamma_0$,
\Omit{
or written explicitly:
\bea
\Gamma_{\gamma N \rightarrow \Delta}^{\alpha \nu}(p,q) 
\equiv i V^{\alpha \nu}_{\Delta out}(p,q) &=& i \frac{e F_\Delta(q^2)}{2 M_{\Delta}^2}
T_3^\dagger\, \gamma_5 \bigg\{
g_1 \left[\, g^{\alpha \nu} \qslash{q} \vslash{p} - \qslash{q} \gamma^\alpha p^\nu 
- \gamma^\alpha \gamma^\nu (p \cdot q) + \vslash{p} q^\alpha \gamma^\nu \, \right]  \nn \\
&& + g_2 \left[\, q^\alpha p^\nu - g^{\alpha \nu} (p \cdot q)\, \right] \nn \\
&& - (g_3/M_{\Delta}) \left[\,q^2 (\gamma^\alpha p^\nu - g^{\alpha \nu} \vslash{p}) +
(q^\alpha \vslash{p} - \gamma^\alpha (p \cdot q) q^\nu )\, \right] \bigg\} \,,
\eqlab{vert_out}
\eea
}
with $p_\alpha$ and $q_\nu$ the four-momenta of the outgoing $\Delta$ and incoming photon, respectively.
The $\gamma N \Delta$ vertex is orthogonal to the four-momenta of both the photon 
and the $\Delta$:
\beq
q_\nu \Gamma_{\gamma \Delta \rightarrow N}^{\nu \alpha}(p,q) = 0, \;\;\;
p_\alpha \Gamma_{\gamma \Delta \rightarrow N}^{\nu \alpha}(p,q) = 0\,. \eqlab{gauge}
\eeq
\Omit{
\bea
& q_\nu \Gamma_{\gamma \Delta \rightarrow N}^{\nu \alpha}(p,q) =
q_\nu \Gamma_{\gamma N \rightarrow \Delta}^{\alpha \nu}(p,q) = 0\,, &  \eqlab{gauge_phot} \\
& p_\alpha \Gamma_{\gamma \Delta \rightarrow N}^{\nu \alpha}(p,q) =
p_\alpha \Gamma_{\gamma N \rightarrow \Delta}^{\alpha \nu}(p,q) = 0\,. & \eqlab{gauge_del}
\eea
}
The first of these equations ensures the usual electromagnetic gauge invariance of the calculation while the second allows us to use only the physical spin $3/2$ component,  
\beq
S^\Delta_{\alpha \beta}(p) = \frac{-i}{\vslash{p}-M_\Delta +i 0}
{\mathcal{P}}^{3/2}_{\alpha \beta}(p)\,,\;\;
{\mathcal{P}}^{3/2}_{\alpha \beta}(p) = g_{\alpha \beta} - {1 \over 3} \gamma_\alpha
\gamma_\beta- {1 \over 3 p^2} \left( \vslash{p} \gamma_\alpha p_\beta + p_\alpha
\gamma_\beta \vslash{p} \right)\,,
\eqlab{prop}
\eeq
of the Rarita-Schwinger propagator~\cite{Rar41}, the background spin $1/2$ component 
vanishing when contracted with the adjacent $\gamma N \Delta$ vertices~\cite{Pas99}.
At present we do not include a width in the $\Delta$ propagator as
its influence on the unpolarised cross section should be small. 
\Omit{
${\mathcal{P}}^{1/2, 22}_{\alpha \beta}(p),
{\mathcal{P}}^{1/2, 12}_{\alpha \beta}(p)$ and ${\mathcal{P}}^{1/2, 21}_{\alpha \beta}(p)$
\beq
S^{RS}_{\alpha \beta}(p) = \frac{-i}{\vslash{p}-M_\Delta +i 0}
 {\mathcal{P}}^{3/2}_{\alpha \beta}(p) +
\frac{2 i}{3 M_\Delta^2} (\vslash{p} + M_\Delta) {\mathcal{P}}^{1/2, 22}_{\alpha \beta}(p) -
\frac{i}{\sqrt{3} M_\Delta}
\left({\mathcal{P}}^{1/2, 12}_{\alpha \beta}(p) + {\mathcal{P}}^{1/2, 21}_{\alpha \beta}(p) \right)\,,
\eqlab{rarsch}
\eeq
\beq
{\mathcal{P}}^{3/2}_{\alpha \beta}(p) = g_{\alpha \beta} - {1 \over 3} \gamma_\alpha \gamma_\beta
- {1 \over 3 p^2} \left( \vslash{p} \gamma_\alpha p_\beta + p_\alpha \gamma_\beta \vslash{p} \right)\,,
\eqlab{proj32}
\eeq
\beq
{\mathcal{P}}^{1/2, 22}_{\alpha \beta}(p) = \frac{p_\alpha p_\beta}{p^2}\,,\;\;
{\mathcal{P}}^{1/2, 12}_{\alpha \beta}(p) =
\frac{p_\alpha p_\beta - \vslash{p} \gamma_\alpha p_\beta}{\sqrt{3}p^2}\,,\;\;
 {\mathcal{P}}^{1/2, 21}_{\alpha \beta}(p) =
\frac{\vslash{p} p_\alpha \gamma_\beta - p_\alpha p_\beta}{\sqrt{3}p^2}\,,
\eqlab{proj12}
\eeq
Also, in this letter we do not include a Coulomb component
in the $\gamma N \Delta$ vertex; the relatively simple vertex~\eqref{vert} 
combines effects of the magnetic and electric components and it has the important property of
retaining only the physical spin $3/2$ component of the $\Delta$ propagator.
}


\Omit{
The box and crossed-box two-photon exchange diagrams with
a $\Delta$ particle in the intermediate state are shown in~\figref{boxdel}.
The corresponding loop integrals can be written as
}
The loop integrals corresponding to the box and crossed-box 
diagrams in~\figref{boxdel} can be written as
\beq 
{\mathcal{M}}_\Delta^{\gamma \gamma} = 
-e^4\!\int \! \frac{d^4 k}{(2 \pi)^4}
\frac{N^\Delta_{box}(k)}{D^\Delta_{box}(k)} -
e^4\!\int \! \frac{d^4 k}{(2 \pi)^4} \frac{N^\Delta_{x-box}(k)}{D^\Delta_{x-box}(k)} \,, 
\eqlab{integr}
\eeq
with the numerators and denominators given by
\bea
N^\Delta_{box}(k) &=& 
\oln{U}(p_4) V_{\Delta in}^{\mu \alpha}(p_2+k,q-k) 
\left[ \vslash{p}_2+\vslash{k}+M_\Delta \right] {\mathcal{P}}^{3/2}_{\alpha \beta}(p_2+k)
V_{\Delta out}^{\beta \nu}(p_2+k,k) U(p_2) \nn \\
& \times &
\oln{u}(p_3) \gamma_\mu \left[ \vslash{p}_1 -\vslash{k} +m_e \right] \gamma_\nu u(p_1)\,, \eqlab{num_box} \\
N^\Delta_{x-box}(k) &=& 
\oln{U}(p_4) V_{\Delta in}^{\mu \alpha}(p_2+k,q-k)
\left[ \vslash{p}_2+\vslash{k}+M_\Delta \right] {\mathcal{P}}^{3/2}_{\alpha \beta}(p_2+k)
V_{\Delta out}^{\beta \nu}(p_2+k,k) U(p_2) \nn \\
& \times &
\oln{u}(p_3) \gamma_\nu \left[ \vslash{p}_3 +\vslash{k} +m_e \right] \gamma_\mu u(p_1)\,, \eqlab{num_x-box} 
\eea
\bea
D^\Delta_{box}(k) & = & \left[ k^2 +i 0 \right] \left[(k-q)^2 +i 0 \right]
\left[ (p_1-k)^2-m_e^2 +i 0 \right] \left[ (p_2+k)^2-M_\Delta^2 + i 0 \right] , \eqlab{den_box} \\
D^\Delta_{x-box}(k) & = & \left. D^\Delta_{box}(k) \right|_{p_1-k \rightarrow p_3+k}\,, \eqlab{den_x-box} 
\eea
where $U$ and $u$ denote the proton and electron four-spinor wave functions, respectively.
Compared to the case of the nucleon~\cite{Blu03},
the presence of a $\Delta$ in the intermediate state entails a more complicated structure of the numerator. 
Also the loop integrals with a $\Delta$ are not infrared divergent, 
in contrast with the nucleon contribution where the infrared part is very important~\cite{Tsa61,Max00}.
The evaluation of~\eqref{integr} involves preliminary algebraic
manipulations to effect cancellations between
terms in the numerators and denominators and subsequent integration of the thus simplified expressions.
The result is obtained analytically in terms of the standard Passarino-Veltman 
dilogarithm functions~\cite{tHo79}.
In the calculation we used the computer package ``FeynCalc"~\cite{Mer91}.

The first and second loop integrals in~\eqref{integr} must be mutually related by crossing symmetry, 
which can be formulated 
in terms of the numerator of~\eqref{del_del}
using the Mandelstam variables $s=(p_1+p_2)^2$,  $t=(p_1-p_3)^2$
and $u=(p_2-p_3)^2 = 2 M_N^2+2 m_e^2 - t - s$.
Denoting $f^{\gamma \gamma}(s,t) \equiv {\mathcal{M}}_B^\dagger \, {\mathcal{M}}_\Delta^{\gamma \gamma}$ 
and writing it as the sum
$f^{\gamma \gamma}(s,t) = f^{\gamma \gamma}_{box}(s,t) + f^{\gamma \gamma}_{x-box}(s,t)$,
where the first (second) term is calculated using only the first (second) integral in~\eqref{integr}, 
the crossing symmetry requires that 
\beq
f^{\gamma \gamma}_{x-box}(s,t) = - \left. f^{\gamma \gamma}_{box}(u,t) \right|_{u=2 M_N^2+2 m_e^2 - t - s} 
\; \Leftrightarrow \;
f^{\gamma \gamma}(s,t) = -f^{\gamma \gamma}(2 M_N^2+2 m_e^2 - t - s, t) \,.
\eqlab{cross}
\eeq
We calculated the integrals in~\eqref{integr} explicitly and 
checked that our results obey the crossing symmetry constraint~\eqref{cross}.


The $\Delta$ two-photon exchange correction to the differential cross section
can be expressed as a quadratic form in the $\gamma N \Delta$ coupling constants 
$g_M = g_1$, $g_E = g_2 - g_1$ and $g_C = g_3$:
\beq
\delta_\Delta = C_{M} \, g_M^2 + C_{ME} \, g_M g_E + C_{E} \, g_E^2
+ C_{C} \, g_C^2 + C_{EC} \, g_E g_C + C_{MC} \, g_M g_C  \,,
\eqlab{delta}
\eeq
with the coefficients depending on the kinematical variables.
The relative contributions of the coupling constants $g_M$, $g_E$ and $g_C$ to $\delta_\Delta$ can be
assessed from~\tabref{coeff}, where the 
$C_{M}$, $C_{ME}$, etc.~are given as functions of $\epsilon$ 
at two fixed $Q^2$ values.
\begin{table}[!htb]
\caption[t1]{The $\epsilon$ dependence of the coefficients 
$C_{M, ME, E, C}$
defined in~\eqref{delta} ($C_{EC, MC} < 10^{-10}$ for any kinematics considered).
The $\gamma N \Delta$ form factor~\eqref{ff} was used with $\Lambda_\Delta=0.84$ GeV.}
\begin{center}
\begin{tabular}{c|cccc|cccc}
 &  & $Q^2=3$ GeV$^2$ & & & & $Q^2=6$ GeV$^2$  & &     \\
\hline
$\epsilon$ & 
$C_M \times 10^4$ & $C_{ME} \times 10^4$ & $C_E \times 10^4$ & $C_C \times 10^4$ &
$C_M \times 10^4$ & $C_{ME} \times 10^4$ & $C_E \times 10^4$ & $C_C \times 10^4$  \\
\hline
0.1  & 2.92  &  1.49 & -1.64 & -1.09  &  3.95 &  3.54 & -5.98 & -5.58  \\
0.2  & 2.53  &  0.94 & -1.61 & -1.00  &  2.07 &  1.72 & -5.74 & -4.98  \\
0.3  & 2.17  &  0.50 & -1.57 & -0.88  &  0.69 &  0.49 & -5.45 & -4.29  \\
0.4  & 1.83  &  0.14 & -1.52 & -0.72  & -0.21 & -0.22 & -5.11 & -3.48  \\
0.5  & 1.54  & -0.11 & -1.45 & -0.50  & -0.81 & -0.63 & -4.72 & -2.52  \\
0.6  & 1.23  & -0.32 & -1.37 & -0.21  & -1.18 & -0.85 & -4.25 & -1.35  \\
0.7  & 0.95  & -0.46 & -1.27 &  0.18  & -1.35 & -0.89 & -3.69 &  0.16  \\
0.8  & 0.65  & -0.55 & -1.15 &  0.79  & -1.31 & -0.77 & -2.99 &  2.33  \\
0.9  & 0.31  & -0.57 & -0.98 &  1.98  & -0.94 & -0.42 & -1.99 &  6.38  \\
\hline
\end{tabular}
\end{center}
\tablab{coeff}
\end{table}
In this calculation we used
the dipole $\gamma N \Delta$ form factor~\eqref{ff} with the cutoff
$\Lambda_\Delta = 0.84$ GeV, which describes a $\Delta$ resonance whose mean-square radius is comparable to that of the nucleon.
This choice is consistent with various 
parametrisations from pion electroproduction~\cite{Cam69,Sat01}.

In the following we will discuss the results obtained with the fixed coupling constants  
$g_M=7$ and $g_E=2$. 
These couplings were used in the Dressed K-matrix Model~\cite{Kon01} 
(adjusted for a different normalisation of the vertex used in the present calculation),
yielding a good coupled-channel description of pion-nucleon scattering, pion photoproduction and 
Compton scattering at low and intermediate energies. For example, the $E2/M1$ ratio obtained in Ref.~\cite{Kon01}
from the pion photoproduction multipoles at the position of the $\Delta$ resonance, is
$R_{EM} = \mbox{Im} E^{3/2}_{1+}/\mbox{Im} M^{3/2}_{1+} \times 100\% \approx -3\%$, 
in agreement with the PDG~\cite{Eid04} value: $-(2.5 \pm 0.5) \%$.
Recent analyses~\cite{Sat01} of pion electroproduction
suggest that the Coulomb coupling constant $g_C$ is small and negative. 
In our calculation we will vary $g_C$ in the range $[-2, 0]$.   
With these values of $g_M$, $g_E$ and $g_C$ one can see 
from~\eqref{delta} and~\tabref{coeff} that the magnetic coupling dominates
the $\Delta$ two-photon exchange correction whereas
the electric coupling has a much smaller effect.
Since the contribution of the Coulomb component is strongly
suppressed (not exceeding $0.2\%$) we will omit it from further discussion, setting $g_C=0$ in the rest of the paper.

The $\epsilon$ dependence of the sum of the $\Delta$ and nucleon two-photon exchange corrections
is shown in~\figref{deleps}, for two
fixed values of $Q^2$. The dependence on the $\gamma N \Delta$ form factor
can be seen by comparing the results obtained with 
the cutoffs $\Lambda_\Delta=0.84$ GeV and $\Lambda_\Delta=0.68$ GeV (the latter choice corresponds to a
$\Delta$ which is spatially ``bigger" than the nucleon).  
\begin{figure}[!htb]
\centerline{{\epsfxsize 14.0cm \epsffile[15 290 570 715]{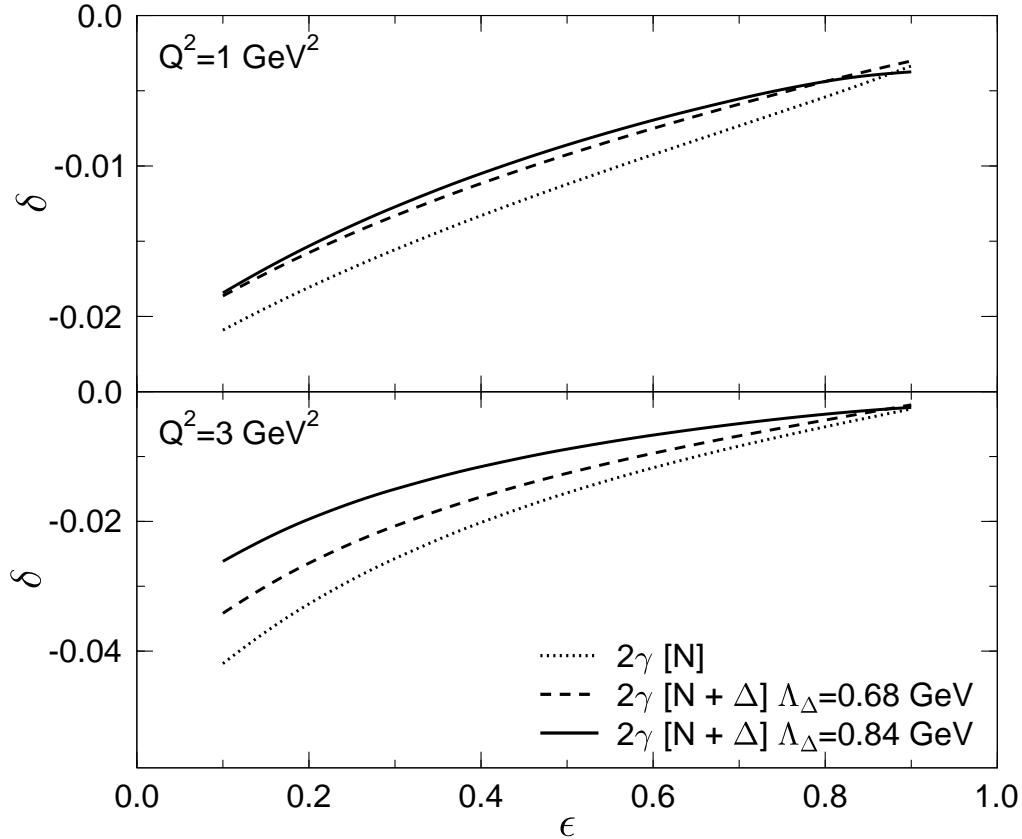}}}
\caption[f2]{Sum of the nucleon (N) and $\Delta$ contributions to the two-photon exchange correction to the
electron-proton scattering cross section, 
using two values of the cutoff $\Lambda_\Delta$. 
\figlab{deleps}}
\end{figure}
The purely nucleon contribution, shown for comparison, was calculated as in
Ref.~\cite{Blu03} using the $\gamma N N$ form factors extracted
from the PT experiments~\cite{Jon00,Bra02}.
The $\Delta$ correction is more prominent at higher momentum transfers.
The $\Delta$ tends to reduce the effect
of the nucleon two-photon exchange, making the modulus of the negative nucleon correction 
somewhat smaller at backward angles (i.~e.~at
low $\epsilon$). 
The combined effect of the nucleon and $\Delta$ two-photon exchanges 
produces a negative correction to the cross section 
at small $\epsilon$, decreasing in magnitude as $\epsilon$ increases.\footnote{The diminishing of the two-photon exchange correction at forward angles 
is consistent with the analysis of electron-proton and positron-proton 
scattering data~\cite{Arr04}.}
The main features of the $\Delta$ contribution -- its smallness and
its tendency to attenuate the nucleon contribution at backward angles -- 
are insensitive to the $\gamma N \Delta$ form factor, being to that extent model-independent.
The detailed interplay between the $\Delta$ and the nucleon contributions
can be more complicated, especially at forward angles, as can be seen from~\figref{deleps}.

The calculated differential cross section is shown by the solid lines in~\figref{csred},
including the Born term and the sum of the two-photon exchange corrections 
$\delta_N + \delta_\Delta$ with the nucleon and the $\Delta$ intermediate states.
\begin{figure}[!htb]
\centerline{{\epsfxsize 14.0cm \epsffile[15 60 570 425]{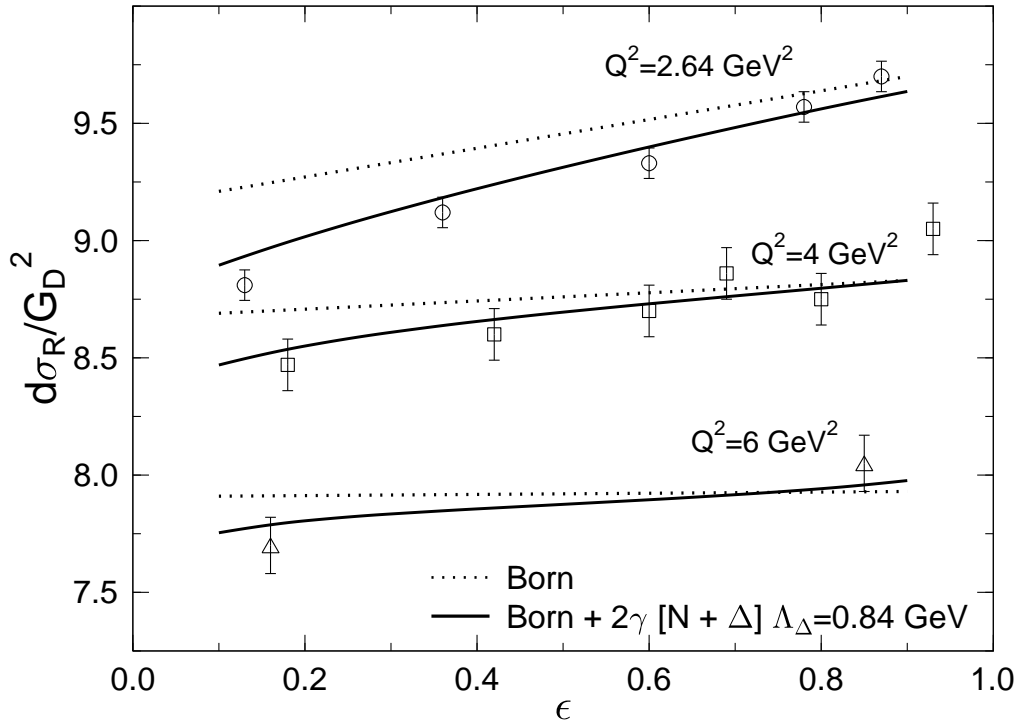}}}
\caption[f3]{Effect of adding the two-photon exchange (with the indicated choice of the
$\gamma N \Delta$ form factor) to the Born cross section, the latter evaluated with the
nucleon form factors from the PT experiment~\cite{Jon00,Bra02}. The reduced cross section is scaled as described in the text. 
The curves for $Q^2=2.64$, $4$ and $6$ GeV$^2$ have been shifted vertically by 
$-0.04$, $+0.04$ and $+0.09$, respectively. 
The data points at three fixed momentum transfers are taken from Refs.~\cite{Wal94,Qat05}.
\figlab{csred}}
\end{figure}
The reduced cross section~\eqref{xsec_red},
scaled for convenience by the square of the standard dipole form factor 
$G_D(Q^2)=1/(1+Q^2/0.84^2)^2$, is compared in~\figref{csred} with the
LT separation measurements from 
SLAC~\cite{Wal94} (at $Q^2=4$ and $6$ GeV$^2$) and JLab~\cite{Qat05} (at $Q^2=2.64$ GeV$^2$).
The dotted lines show the Born contribution alone, using the nucleon form factors $G_{E,M}(Q^2)$ 
taken from the analysis of the JLab PT experiment~\cite{Jon00,Bra02}. 
One can see that including only the Born term is inadequate in the analysis of the data.
The addition of the two-photon exchange correction increases the slope of the cross section,
also exhibiting some nonlinearity in $\epsilon$.  
Thus the results of the PT and LT separation experiments
become essentially compatible by including the nucleon and $\Delta$ 
two-photon exchange corrections. 

To summarise,
we calculated the correction to the electron-proton scattering cross section due to
the two-photon exchange with a $\Delta$ intermediate state, treated on the same 
footing as the intermediate nucleon contribution.
For realistic choices of the $\gamma N \Delta$ vertex
we found that the $\Delta$ contribution alters the cross section 
by an amount from $-1\%$ to $+2\%$, and is largest at backward scattering angles. 
For the cross section obtained using the LT separation technique, the
$\Delta$ two-photon exchange contribution
slightly reduces the magnitude of the (negative) nucleon correction.
Generally, the cross section including the nucleon and $\Delta$ 
two-photon exchange corrections
has the angular dependence which can accommodate the results of both
the LT separation and PT methods of measuring the nucleon form factors.
This calculation therefore provides explicit and compelling evidence that the
two-photon exchange contribution (with the lowest mass, $N$ and $\Delta$
intermediate states) can resolve the form factor discrepancy.
To reconcile these two methods completely,
theoretical analyses of the data might need additional ingredients.
For example, one may take into account the
dependence of the $\gamma N N$ and $\gamma N \Delta$ vertices on the nucleon 
and $\Delta$ off-shell momenta (as was suggested in~\cite{Kon05}).
Heavier hadron resonances or quark degrees of freedom
should also become important at higher momentum transfers (see e.~g.~\cite{Afa05}).

\begin{acknowledgments}

This work of S.~K.~and P.~G.~B.~was supported in part by NSERC (Canada).
The Southeastern Universities Research Association (SURA) operates the Thomas Jefferson National Accelerator Facility (Jefferson Lab) for the DOE under contract
DE-AC05-84ER-40150. We thank John Arrington for useful comments.
\end{acknowledgments}



\begin{thebibliography}{99}
\bibitem{Wal94} R.~C.~Walker et al., \PRD{49}{1994}{5671}; 
                               L.~Andivahis et al., \PRD{50}{1994}{5491}.
\bibitem{Qat05} I.~A.~Qattan et al., \PRL{94}{2005}{142301}.
\bibitem{Jon00} M.~K.~Jones et al., \PRL{84}{2000}{1398}; 
                O.~Gayou et al., \PRL{88}{2002}{092301}.
\bibitem{Bra02} E.~J.~Brash, A.~Kozlov, Sh.~Li, and G.~M.~Huber, \PRC{65}{2002}{051001(R)}.
\bibitem{Arr03} J.~Arrington, \PRC{68}{2003}{034325}.
\bibitem{Blu03} P.~G.~Blunden, W.~Melnitchouk, and J.~A.~Tjon, \PRL{91}{2003}{142304};
                P.~G.~Blunden, W.~Melnitchouk, and J.~A.~Tjon, nucl-th/0506039. 
\bibitem{Gui03} P.~A.~M.~Guichon and M.~Vanderhaegen, \PRL{91}{2003}{142303};
                              Y.~C.~Chen, A.~Afanasev, S.~J.~Brodsky, C.~E.~Carlson, 
                               and M.~Vanderhaegen, \PRL{93}{2004}{122301}.
\bibitem{Dre57} S.~D.~Drell and M.~Ruderman, \PR{106}{1957}{561}; S.~D.~Drell and S.~Fubini, \PR{113}{1959}{741};
                               G.~K.~Greenhut, \PR{184}{1969}{1860}.
\bibitem{Cam69} J.~A.~Campbell, \PR{180}{1969}{1541}.
\bibitem{Max00} L.~C.~Maximon and J.~A.~Tjon, \PRC{62}{2000}{054320}.
\bibitem{Bjo64} J.~D.~Bjorken and S.~D.~Drell,
                {\it Relativistic Quantum Mechanics} (McGraw-Hill, 1964).               
\bibitem{Kon01} S.~Kondratyuk and O.~Scholten, \PRC{64}{2001}{024005}.
\bibitem{Rar41} W.~Rarita and J.~Schwinger, \PR{60}{1941}{61}.
\bibitem{Pas99} V.~Pascalutsa and R.~G.~E.~Timmermans, \PRC{60}{1999}{042201}.
\bibitem{Tsa61} Y.~S.~Tsai, \PR{122}{1961}{1898}.
\bibitem{tHo79} G.~'t Hooft and M.~Veltman, \NPB{153}{1979}{365}; 
                               G.~Passarino and M.~Veltman, \NPB{160}{1979}{151}. 
\bibitem{Mer91} R.~Mertig, M.~B{\"o}hm, A.~Denner, Comp.~Phys.~Comm.~{\bf 64}, 345 (1991)
                               (available at http://www.feyncalc.org/).
\bibitem{Eid04} S.~Eidelman et al. (Particle Data Group), \PLB{592}{2004}{1}.
\bibitem{Arr04} J.~Arrington, \PRC{69}{2004}{032201}.
\bibitem{Sat01} T.~Sato and T.-S.~H.~Lee, \PRC{63}{2001}{055201};
                G.~L.~Caia, V.~Pascalutsa, J.~A.~Tjon, and L.~E.~Wright, \PRC{70}{2004}{032201}.
\bibitem{Kon05} S.~Kondratyuk, K.~Kubodera, and F.~Myhrer, \PRC{71}{2005}{028201}.
\bibitem{Afa05} A.~V.~Afanasev, S.~J.~Brodsky, C.~E.~Carlson, Y.-C.~Chen, and M.~Vanderhaegen, hep-ph/0502013.
\end{thebibliography}
\end{document}